\documentclass{aa}
\usepackage{graphicx}
\usepackage{natbib}
\bibpunct{(}{)}{;}{a}{}{,}
\begin{document} 

   \title{The contribution of galaxies to the UV ionising background \\
	  and the evolution of the Lyman forest}
 
   \author{ Simone Bianchi\inst{1}\fnmsep\thanks{Fellow of the European
                                          Community Research and Training 
					  Network: \emph{ The Physics of 
					  the Intergalactic Medium}.}
	    \and 
	    Stefano Cristiani\inst{2}\fnmsep\inst{3} 
	    \and 
	    Tae-Sun Kim\inst{1}
	  }
 
   \offprints{sbianchi@eso.org}
 
   \institute{ 
   European Southern Observatory, Karl-Schwarzschild-Strasse 2, D-85748 
   Garching, Germany
   \and
   Space Telescope European Coordinating Facility, ESO,
   Karl-Schwarzschild-Strasse 2, D-85748 Garching, Germany
   \and
   Dipartimento di Astronomia dell'Universit\`a di Padova, Vicolo
   dell'Osservatorio, I-35122 Padova, Italy
   } 
 
   \date{Received  / Accepted }

\abstract{
We have modelled the evolution of the number of Ly$\alpha$ absorbers with 
redshift, resulting from the evolution of the ionising background and the 
Hubble expansion. The contribution of quasars (QSOs) and galaxies to the 
\ion{H}{i}-ionising UV background has been estimated. The QSOs emissivity 
is derived from recent fits of their luminosity function. The galaxy emissivity 
is computed using a stellar population synthesis model, with a 
star-formation history scaled on observations of faint galaxies at 
$\lambda\ge 1500$\AA. We allow for three values of the fraction of ionising 
photons that can escape the interstellar medium, $f_\mathrm{esc}=0.05$,
0.1 and 0.4. The Intergalactic Medium is modelled as made of purely-absorbing 
clouds with the distribution in redshift and column density obtained 
from QSOs absorption lines. For the adopted values of $f_\mathrm{esc}$,
the contribution of galaxies to the ionising UV background is comparable or
greater than that of QSOs. Accounting for the contribution of clouds to the 
UV emission, all models with $f_\mathrm{esc}\la 0.1$ provide an ionising flux 
compatible with local and high-$z$ determination, including those with a 
pure QSOs background. The observed $z\sim 1$ break in 
the evolution can be better explained by a dominant contribution from 
galaxies. We find that models in $\Lambda$-cosmology with
$\Omega_\mathrm{m}$=0.3, $\Omega_\Lambda$=0.7 describe the 
flat absorbers evolution for $z\la 1.0$ better than models for
$\Omega_\mathrm{m}$=1.0.
}

\titlerunning{The contribution of galaxies to the UV ionising background}
\maketitle 

\section{Introduction}

The evolution of the Lyman forest is governed by two main factors: the
Hubble expansion and the the metagalactic UV background.
At high redshift the expansion, which tends to increase the ionisation
of the matter, and the UV background, increasing or non-decreasing 
with decreasing redshift, work in the same direction and
cause a steep evolution of the number of lines with $z$.
At low redshift, the UV background starts to decrease with decreasing
redshift, due to the reduced number of ionising sources, and this
effect counteracts the Hubble expansion. As a result the evolution 
of the number of lines slows down.

In a recent study of the evolution of the Lyman forest
\citet*{KimA&A2001} have shown that the number of Ly$\alpha$ lines per unit
redshift, $dN/dz$, is well described by a double power-law with a
break at $z \sim 1$. For column densities in the interval
$N_\ion{H}{i} = 10^{13.64-16} \ \mathrm{cm}^{-2}$, $dN/dz \propto
(1+z)^{2.19 \pm 0.27}$ at $1.5 <z < 4$, while $dN/dz
\propto (1+z)^{0.16 \pm 0.16}$ at $z<1$ \citep{WeymannApJ1998}.

Recent numerical simulations have been remarkably successful in
reproducing the observed evolution 
\citep[see, for example ][]{DaveApJ1999,MachacekApJ2000},
leaving little doubt about the general interpretation of the phenomenon. 
However the same simulations predict the break in the $dN/dz$ 
power-law at a redshift $z \sim 1.8$ which appears too high in the
light of the new results of \citet{KimA&A2001}.
This suggests that the UV background implemented in the simulations is
not completely correct: QSOs have been considered as the main source
of ionising photons, and, since their space density drops below $z\sim
2$, so does the UV background.
However, galaxies can produce a non-negligible ionising flux too, perhaps
more significant than previously assumed, as shown by recent measurements
by \citet*{SteidelApJ2001}. The galaxy contribution can keep the
UV background relatively high until at $z \sim 1$ the global star
formation rate in the Universe quickly decreases, determining the
change in the number density of lines.

In this paper we recompute the contribution of QSOs and galaxies to 
the UV background following the recipes of
\citet{MadauApJL1991,MadauApJL1992,HaardtApJ1996} and \citet*{MadauApJ1998} 
(Sect.~\ref{method}).  In Sect.~\ref{results} the results are compared with 
constraints on the UV background derived from the proximity effect and the 
H$\alpha$ emission of high galactic latitude clouds. In Sect.~\ref{results2} 
the evolution of the number of Ly$\alpha$ lines per unit redshift is computed 
according to a simple analytical model for different relative contributions 
of galaxies and QSOs. In Sect.~\ref{conclu} we summarise the results and 
discuss some consequences of them in terms of present and future observations.

All the computations are carried out for two flat cosmologies:
an Einstein-De Sitter cosmology
($\Omega_m=1,\Omega_\Lambda=0$) and a $\Lambda$ cosmology
($\Omega_m=0.3,\Omega_\Lambda=0.7$). We have adopted 
$H_0$=70 km s$^{-1}$ Mpc$^{-1}$ \citep{FreedmanApJ2001} throughout
and scaled to this value the data derived from the literature.

\section{The Ultraviolet Background}
\label{method}

The mean specific intensity of the Ultraviolet background
$J(\nu_\mathrm{obs},z_\mathrm{obs})$, as seen at a frequency
$\nu_\mathrm{obs}$ by an observer at redshift $z_\mathrm{obs}$, can be
derived from
\begin{eqnarray}
J(\nu_\mathrm{obs},z_\mathrm{obs})=&\frac{1}{4\pi}&
\int_{z_\mathrm{obs}}^{\infty}
{(1+z_\mathrm{obs})^3 \over (1+z)^3} \epsilon(\nu,z) \nonumber
\\
&\times& \;\; e^{-\tau_\mathrm{eff}(\nu_\mathrm{obs},z_\mathrm{obs},z)}
\frac{dl}{dz} dz,
\label{transfer}
\end{eqnarray}
where $\nu=\nu_\mathrm{obs}(1+z)/(1+z_\mathrm{obs})$, $\epsilon(\nu,z)$
is the proper space-averaged volume emissivity, 
$\tau_\mathrm{eff}(\nu_\mathrm{obs},z_\mathrm{obs},z)$ is
the effective optical depth at $\nu_\mathrm{obs}$ of the Intergalactic
Medium (IGM) between redshifts $z_\mathrm{obs}$ and $z$, and $dl/dz$
the proper line element \citep{MadauApJL1991,MadauApJL1992,HaardtApJ1996}.

The emissivity $\epsilon(\nu,z)$ should include a contribution both from 
direct sources of UV radiation (e.g. QSOs and galaxies) and from the IGM
clouds themselves, through continuum radiative recombination of the gas
\citep{HaardtApJ1996}. For the sake of simplicity, we consider here the
case of a purely absorbing IGM, thus omitting radiative recombination.
The effect of this omission will be discussed later.

The line element can be written as
\begin{equation}
\frac{dl}{dz}=\frac{c}{H(z) (1+z)},
\label{dldz}
\end{equation}
where $c$ is the velocity of light and
$H(z)=H_0 [\Omega_m (1+z)^3 + \Omega_\Lambda]^{1/2}$ 
is the Hubble parameter for a flat universe 
($\Omega=\Omega_m+\Omega_\Lambda=1$).

\subsection{Opacity}
\label{opacity}
The effective optical depth $\tau_\mathrm{eff}$ through the IGM is defined 
as $e^{-\tau_\mathrm{eff}}=\langle e^{-\tau} \rangle$, where the mean is 
taken over all the lines of sight from the redshift of interest. For a 
Poisson distribution of discrete absorbers
\citep{ParesceApJ1980,MollerA&A1990,MadauApJL1991,MadauApJL1992},
\begin{eqnarray}
\tau_\mathrm{eff}(\lambda_\mathrm{obs},z_\mathrm{obs},z)&=& \nonumber \\
\int_{z_\mathrm{obs}}^z \!\!\!\!\! &dz'&\int_0^{\infty} 
\!\!\!\! dN_\ion{H}{i} 
\; f(N_\ion{H}{i},z')
(1-e^{-\tau(\lambda')}),
\label{taueff} 
\end{eqnarray}
where $f(N_\ion{H}{i},z')=
\partial^2 N / \partial N_\ion{H}{i}\partial z'$ is the
distribution of absorbers as a function of redshift and column 
density of the atomic hydrogen $N_\ion{H}{i}$, 
and $\tau(\lambda')$ is the optical depth of an individual cloud 
for ionising radiation at a wavelength 
$\lambda'=\lambda_\mathrm{obs}(1+z_\mathrm{obs})/(1+z)$.
For 228\AA$<\lambda'\leq$912\AA, the main contribution to absorption of
UV photons is given by the ionisation of \ion{H}{i}, therefore
\begin{equation}
\tau(\lambda')=
\tau_\ion{H}{i} (\lambda')=
N_\ion{H}{i}\, \sigma_\ion{H}{i} \, 
\left(\frac{\lambda'}{912 \mbox{\AA}}\right)^3,
\end{equation}
where $\sigma_\ion{H}{i}=6.3 \cdot 10^{-18}$ cm$^{-2}$ is the 
photo-ionisation cross-section at the Lyman limit for \ion{H}{i}
\citep{OsterbrockBook1989}. 

The ionisation of \ion{He}{i} at 504\AA\ is not considered: \ion{He}{i}
being almost completely ionised, its contribution to the total opacity is 
negligible \citep{HaardtApJ1996}. For completeness, we have included the
contribution of \ion{He}{ii} ionisation to the opacity, although it
does not affect the results presented in this paper.
\ion{He}{ii} is ionised for $\lambda'\leq$228\AA: thus,
\begin{equation}
\tau(\lambda')=
\tau_\ion{H}{i}+
N_\ion{He}{ii}\, \sigma_\ion{He}{ii} \, 
\left(\frac{\lambda'}{228 \mbox{\AA}}\right)^3,
\end{equation}
with $\sigma_\ion{He}{ii}=1.58 \cdot 10^{-18}$ cm$^{-2}$ 
\citep{OsterbrockBook1989}. The column density $N_\ion{He}{ii}$ can be
derived from $N_\ion{H}{i}$ by solving the radiative transfer within a 
cloud. When clouds are optically thin at 228\AA, a simple solution can
be found \citep{HaardtApJ1996}: 
\begin{equation}
N_\ion{He}{ii}\,\approx\, 1.8\, N_\ion{H}{i}\, 
\frac{J(912\mbox{\AA})}{J(228\mbox{\AA})}.
\label{ratio}
\end{equation}
While Eqn.~(\ref{ratio}) does not hold for optically thick clouds, its use
in the optically thick case can still provide correct estimates for
the cosmic opacity, when $\lambda_\mathrm{obs} > 228$\AA\ 
\citep{HaardtApJ1996}. Since the \ion{He}{ii} contribution to
$\tau_\mathrm{eff}$ depends on the UV background, we have solved
Eqn.~(\ref{transfer}) iteratively for every value of $z_\mathrm{obs}$.

For the redshift and column density distribution of absorption lines we
have adopted the usual form
\begin{equation}
f(N_\ion{H}{i},z)= \left(\frac{A}{10^{17}}\right)
\left(\frac{N_\ion{H}{i}}{10^{17}\mathrm{\;\;cm^{-2}}}\right)^{-\beta} 
(1+z)^\gamma.
\label{absdistr}
\end{equation}
Fits of the absorption line  distribution show that the index $\beta$
varies for different ranges of $N_\ion{H}{i}$ \citep{FardalAJ1998,KimA&A2001}.
However, it is possible to describe the cloud distribution with $\beta=1.46$ 
over several decades in $N_\ion{H}{i}$ \citep{PetitjeanMNRAS1993}. For 
simplicity, we have adopted this single value over the whole column
density range considered here. \citet{KimA&A2001} combined high resolution
VLT/UVES observations of 3 QSOs with literature data and derived a
line number density per unit redshift $dN/dz=9.06 (1+z)^{2.19}$, for
the column density range $N_\ion{H}{i}=10^{13.64-16} \mathrm{cm^{-2}}$
and $z>1.5$. At lower redshifts, HST observations from the QSO absorption 
line key project show a slower evolution with redshift, with $dN/dz=34.7\;
(1+z)^{0.16}$ \citep{WeymannApJ1998}. The change in evolution, according 
to the results of \citet{WeymannApJ1998} and \citet{KimA&A2001}, occurs at 
$z\sim 1$. In this paper, we use $\gamma$=2.19 to describe the redshift
evolution of the Ly$\alpha$ forest ($10^{13} \leq N_\ion{H}{i}/
\mathrm{cm^{-2}} \leq 1.58\cdot10^{17}$) at $z>1$, and $\gamma$=0.16 at
$z\leq 1$. By choosing $A=0.13$ and $A=0.50$, the integral of 
Eqn.~(\ref{absdistr}) over $N_\ion{H}{i}=10^{13.64-16} \mathrm{cm^{-2}}$ 
reproduces the results of \citet{KimA&A2001} and \citet{WeymannApJ1998},
respectively. The distribution of Lyman Limit systems is derived in an
analogous way from \citet{StorrieLombardiApJL1994}. For $1.58\cdot 10^{17} 
\leq N_\ion{H}{i}/ \mathrm{cm^{-2}} \leq 10^{20}$, we use $A=0.17$ and 
$\gamma$=1.55. The parameters adopted for the distribution of absorbers
are summarized in Tab.~\ref{partable}.

\begin{table}
\begin{center}
\begin{tabular}{lllll}
$A$ &  $\beta$  &  $\gamma$ & $N_\ion{H}{i}/\mathrm{cm^{-2}} $ & \\
\hline
0.50  &  1.46  &  0.16  & $10^{13} - 1.58\cdot 10^{17}$ & $z\leq 1$ \\
0.13  &  1.46  &  2.19  & $10^{13} - 1.58\cdot 10^{17}$ & $z> 1$    \\
0.17  &  1.46  &  1.55  & $1.58\cdot 10^{17} - 10^{20}$ &  \\
\hline
\end{tabular}
\end{center}
\caption{Parameters adopted for the distribution of absorbing clouds
(Eqn.~\ref{absdistr}).}
\label{partable}
\end{table}

\begin{figure}[b]
\resizebox{\hsize}{!}{\includegraphics{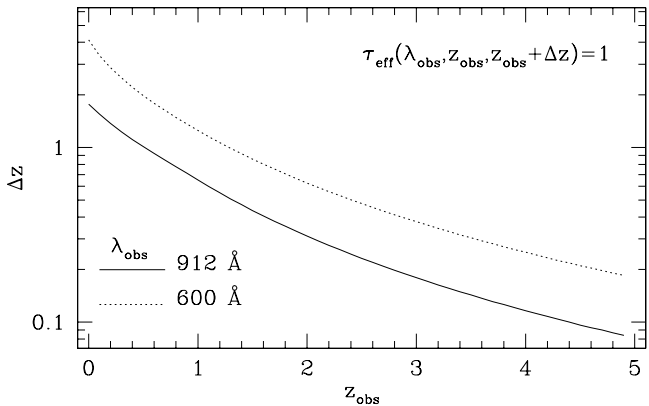}}
\caption{Distance in redshift $\Delta z$ between an observer at
$z_\mathrm{obs}$ and a point at $z=z_\mathrm{obs}+\Delta z$ for which
$\tau_\mathrm{eff}(\lambda_\mathrm{obs},z_\mathrm{obs},z)=1$, as a
function of $z_\mathrm{obs}$. Plots are shown for radiation observed 
at the Lyman edge and for $\lambda_\mathrm{obs} = 600$\AA.
The plot does not include the contribution to the opacity of the
\ion{He}{ii} ionisation, that will affect the results at 600\AA. 
Assuming $J(912\mbox{\AA})/J(228\mbox{\AA})$=50 (the minimum
value in our simulations, when only QSOs contribute to the background),
$\Delta z$ at 600\AA\ will decrease to $\approx 2$ in the range $0<z<0.5$.}
\label{zreach}
\end{figure}

As already pointed out by many authors
\citep{MadauApJL1991,MadauApJL1992,HaardtApJ1996},
the UV background becomes more dominated by local sources as the redshift
increases. This can be seen in Fig.~\ref{zreach}, where we show the
distance in redshift $\Delta z=z-z_\mathrm{obs}$ corresponding to an 
effective optical depth
$\tau_\mathrm{eff}(\lambda_\mathrm{obs},z_\mathrm{obs},z)=1$, as a
function of $z_\mathrm{obs}$. For radiation at
$\lambda_\mathrm{obs}=912$\AA\, $\Delta z$ decreases from 1.8 at
$z_\mathrm{obs}$=0 to 0.08 at $z_\mathrm{obs}$=5. The same trend can be
seen for $\lambda_\mathrm{obs}=600$\AA\, but with larger values of
$\Delta z$ (less absorption), because of the
dependence of the \ion{H}{i} ionisation cross-section on the wavelength.
Since only radiation from local sources can easily reach
$z_\mathrm{obs}$, it is not necessary to compute the integral in
Eqn.~(\ref{transfer}) up to $z=\infty$ (or to the maximum $z$ for which
UV emitting sources are available). We have used in our calculation
$z_\mathrm{max}$=5.

The absorber distribution we have adopted produces opacities that are very 
similar to those of \citet{HaardtApJ1996}. \citeauthor{FardalAJ1998} 
(\citeyear{FardalAJ1998}; see also \citealt{ShullAJ1999}) have derived
values for $A$, $\beta$ and $\gamma$ by fitting the distribution of
absorption lines in several ranges of column density. The opacity provided
by their model is smaller than the one presented here. For example, at 
$z_\mathrm{obs}$=3 we reach $\tau_\mathrm{eff}(912$\AA$)=1$ for $\Delta
z=0.18$, while it is $\Delta z=0.25$ for model A2 in \citet{FardalAJ1998}.
For the same emissivities, the opacity of \citet{FardalAJ1998} will
result in a UV background higher than ours. The difference increases
with $z_\mathrm{obs}$ and reaches 0.1 dex at $z_\mathrm{obs}$=3. 

\subsection{QSO emissivity}
\label{qsodis}
The QSO contribution to the UV emissivity has been derived from the QSO
luminosity function, for which we have adopted the double power-law 
Pure Luminosity Evolution model \citep{BoyleMNRAS1988}
\begin{equation}
\phi(L,z)=\frac{\phi^\star}{
L^\star(z)\left[
\left(\frac{L}{L^\star(z)}\right)^{\beta_1}+
\left(\frac{L}{L^\star(z)}\right)^{\beta_2}
\right]},
\label{ple}
\end{equation}
where $\beta_1$ and $\beta_2$ are the faint- and bright-end of the
luminosity function. A few functional forms have been adopted for
the redshift evolution of the break luminosity $L^\star(z)$. 
Using a sample of over 6000 QSOs with $0.35 <z < 2.3$, 
\citet{BoyleMNRAS2000} find that the evolution is well fitted by a 
second-order polynomial of the form
\begin{equation}
L^\star(z)=L^\star(0)\; 10^{k_1 z + k_2 z^2}.
\label{lumevol}
\end{equation}
At redshifts $z\sim 2$, the luminosity evolution stops and the
comoving number density of QSOs remains constant up to $z\sim 3$
\citep{BoyleMNRAS1988}. 
At $z > 3$ the QSO number density declines dramatically. 
A recent study by \citet{FanAJ2001} on a sample of 39 high-redshift
QSOs from the Sloan Digital Sky Survey, suggests that the number density
declines as $e^{-1.15z}$ in the redshift range $3.6 < z < 5$.

\citet{BoyleMNRAS2000} provide the best-fitting parameters of the
B-band luminosity function for the two cosmologies adopted in this
paper. The B-band proper emissivity can then be derived through the 
integral
\begin{equation}
\begin{array}{lcl}
\epsilon(\nu_B,z)&=&
(1+z)^3 \; \left.\int_{L_\mathrm{min}}^\infty\right. \!\!\! L 
\;\phi(L,z) dL \\ \nonumber \\ \label{qsoemi}
&=& (1+z)^{3}\; 10^{k_1 z + k_2 z^2} \; \epsilon(\nu_B,0), \nonumber
\end{array}
\end{equation}
where the factor $(1+z)^3$ is used to transform comoving into proper
number densities. For the flat Einstein-De Sitter universe and the
$\Lambda$-cosmology, we have derived $\epsilon(\nu_B,0)=1.2\cdot 10^{24}$
and $7.1\cdot 10^{23} \;h_{70}\;\mathrm{erg\; s^{-1}\; Hz^{-1}\; Mpc^{-3}}$,
respectively. $L_\mathrm{min}$ scales with $z$ as $L^\star$ and is
chosen to correspond to $M_\mathrm{B}^\mathrm{max}=-22$ at $z=3$, for both
cosmologies
\citep[$M_\mathrm{B}^\mathrm{max}\approx -18$ at $z=0$;][]{MadauApJL1992}. 
Since Eqn.~(\ref{lumevol}) well describes a flat luminosity evolution for 
$z\sim2$, we have adopted the emissivity of Eqn.~(\ref{qsoemi}) up to $z=3$. 
For $z>3$ we have adopted the exponential decline of \citet{FanAJ2001}.

\citet{BoyleMNRAS2000} derived the B-band luminosity function from 
observations in the QSOs UV restframe, applying the K-correction 
for the composite QSO spectrum of \citet{CristianiA&A1990}. For
consistency, we have used the same spectrum to derive the UV emissivity 
for $\lambda >1050$\AA.  For $\lambda< 1050$\AA\ we have used a power-law,
$\epsilon(\nu)\propto \nu^{-1.8}$, as measured on a sample of
radio-quiet QSOs observed with HST \citep{ZhengApJ1997}. 

We have also derived the QSO emissivity from the work of 
\citet{LaFrancaAJ1997}. They fitted the luminosity function on a
sample of 326 objects (the Homogeneous Bright QSO Survey) by using a
different luminosity evolution. Results obtained with this emissivity
are very similar to those for the emissivity discussed above
and are not presented in this paper.

\begin {figure*}[t]
\sidecaption
\includegraphics[width=12cm]{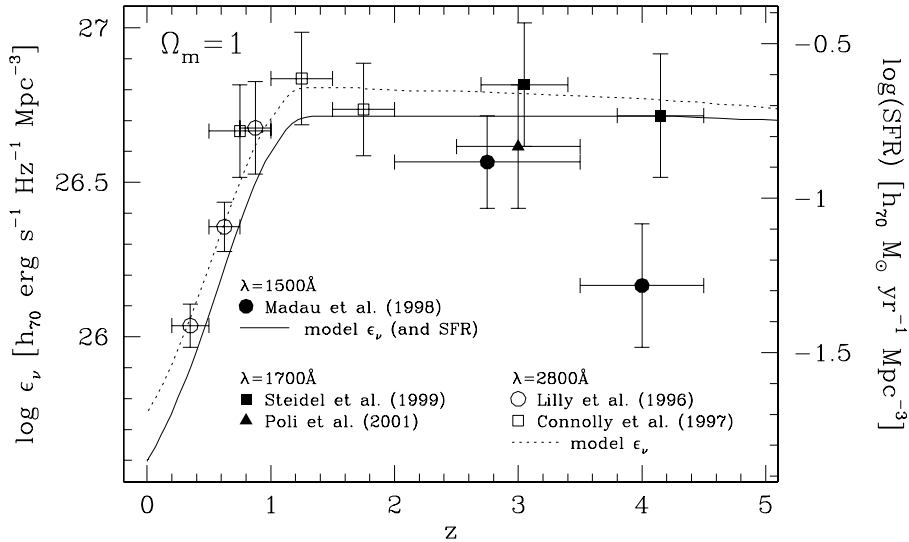}
\caption{Comoving UV emissivity in a flat $\Omega_\mathrm{m}=1$ universe, 
from literature (dots) and from the spectral synthesis model described in 
Sect.~\ref{galemi} (lines). The model emissivity has been corrected for 
internal dust extinction, using the \citeauthor{CalzettiProc1997}'s 
(\citeyear{CalzettiProc1997}) attenuation law with $E(B-V)=0.1$. Because of 
Eqn.~(\ref{normaliz}) and of the redshift-independent dust correction, 
emissivity at $\lambda=1500$\AA\ is proportional to the SFR history adopted 
in the model (scale on right ordinate).
}
\label{sfrh}
\end{figure*}

\subsection{Galaxy emissivity}
\label{galemi}

The galactic emissivity in the ionising UV has been derived following the 
method outlined by \citet{MadauApJ1998}. The comoving UV emissivity at
$\lambda\ge 1500$\AA\ (rest frame) can be derived from galaxy surveys 
as a function of the redshift. Because UV light is mainly produced by
short-lived OB stars, it is possible to convert the UV emissivity into a
star-formation history of the universe. If we assume that the mean 
luminosity evolution of the galaxies in the universe can be described with 
a single galactic spectrum compatible with the derived star-formation 
history, a stellar population synthesis model can be used to derive the 
emissivity at any wavelength.

We have used the latest version of the \citet{BruzualApJ1993} stellar
population synthesis models, updated with a new set of stellar evolutionary
tracks and spectral libraries (\citealt{BruzualPrep}; see also  
\citealt{LiuApJ2000}). A \citet{SalpeterApJ1955} IMF with $0.1 <
M_\star/M_\odot < 100$ has been adopted. We have chosen the quasi-empirical 
library of stellar spectra for solar metallicity, derived from observations 
for $\lambda>1150$\AA\ and from model of stellar atmospheres for
$\lambda<1150$\AA. The conversion factor between UV luminosity and SFR has 
been derived from the adopted stellar model, in the limit of continuous
star-formation \citep{MadauApJ1998,KennicuttARA&A1998}. For radiation
at $\lambda=1500$\AA\, we obtain
\begin{equation}
\frac{L_\mathrm{UV}}{SFR}=7.9\cdot 10^{27}\;\; \frac{\mathrm{erg\; s^{-1 } \;
Hz^{-1}}}{\mathrm{M_\odot\; yr^{-1}}}.
\label{normaliz}
\end{equation}
Conversion factors for other choices of the IMF and stellar libraries among
those provided by the \citeauthor{BruzualPrep} model typically differ over a 
range of $\approx$ 0.3 dex. The same span is also observed when
comparing conversion factors obtained with different models
\citep{KennicuttARA&A1998}.

The star-formation history at $z<2$ has been calibrated on the UV comoving
emissivities tabulated by \citet{MadauApJ1998} for a flat
$\Omega_\mathrm{m}$=1 universe. Emissivities at 
$\lambda=2800$\AA\ have been derived from the Canada-France Redshift 
Survey \citep{LillyApJL1996} in the range $0.2 < z < 1$ and from a HDF-north
sample of objects with optical photometric redshifts for $1<z<2$ 
\citep{ConnollyApJL1997}. \citet{MadauMNRAS1996} have derived the emissivity 
at $\lambda=1500$\AA\ from objects selected in two redshift ranges at 
$z\sim 3$ and $z\sim 4$, from object selected on the HDF-north with the UV 
dropout technique. The lower emissivities at $z> 2$ suggested that the star 
formation rate has reached a maximum at $z\sim 1 - 2$. However, a different 
picture emerged from the ground-based survey of \citet{SteidelApJ1999}. 
They selected Lyman break galaxies on a larger area than the HDF, refining 
the colour selection criteria with spectroscopic observations of a few object 
in the sample. The derived emissivity (at $\lambda=1700$\AA) for $z\sim 3$ is 
still consistent with the \citet{MadauMNRAS1996} value, while the emissivity 
at $z\sim 4$ does not show a steep decline. A value for the emissivity
at $\lambda=1700$\AA\ in the redshift bin $2.5 < z < 3.5$ can be derived 
from the luminosity function fitted by \citet{PoliApJL2001} on a 
combined ground-based and HST database. The datapoints are shown in
Fig.~\ref{sfrh} as a function of redshift. We have corrected the emissivities 
of \citet{SteidelApJ1999} and \citet{PoliApJL2001} to include all objects with
luminosities from 0 to $\infty$. The large errors in the data derived from 
\citet{SteidelApJ1999} 
and \citet{PoliApJL2001} reflect the uncertainties in the faint-end slope 
$\alpha$ of the \citet{SchechterApJ1976} luminosity function. 
\citet{SteidelApJ1999} derive $\alpha=-1.60 \pm 0.13$ and \citet{PoliApJL2001}
$\alpha=-1.37\pm0.20$. In the \citet{MadauApJ1998} tabulation,
$\alpha=-1.3$ was used.

A smooth star-formation history has been derived from the observed UV
emissivities in Fig.~\ref{sfrh}, by using Eqn.~(\ref{normaliz}) and correcting 
for dust internal extinction according to the \citeauthor{CalzettiProc1997}'s
(\citeyear{CalzettiProc1997}) attenuation law. At high redshift, we have
adopted the flat star-formation history suggested by the work 
of \citet{SteidelApJ1999}. Synthetic galactic spectra have then been 
produced with the \citet{BruzualPrep} code. In Fig.~\ref{sfrh} we show the
evolution of the modelled UV emissivity at 1500\AA\ (solid line) and
at 2800\AA\ (dotted line), for a colour excess E(B-V)=0.1. The model is 
compatible with both data at 2800\AA\ and $z<2 $ and data at 1500\AA\ 
(and 1700\AA) and $z>2$. For the chosen E(B-V), the model is also consistent 
with the evolution of the emissivity in the optical-NIR regime for $z<2$,
as tabulated by \citet[][not shown in Fig.~\ref{sfrh}]{MadauApJ1998}.

Unfortunately, it has not been possible to repeat the same procedure to 
model the emissivity (and star-formation history) in $\Lambda$-cosmology. 
Only \citet{PoliApJL2001} present a luminosity function derived assuming
$\Omega_m=0.3,\Omega_\Lambda=0.7$. \citet{SteidelApJ1999} give the 
emissivity for the objects visible in their survey, but they do not
provide a luminosity function for an extrapolation to fainter
luminosities. Therefore, we have used the model of the emissivity for 
$\Omega_\mathrm{m}$=1 and scaled it with a redshift-dependent correction:
for sufficiently small redshift bin, it can be shown that the ratio
of the emissivities in the flat $\Lambda$- and Einstein-De Sitter
cosmologies is $\sqrt{0.7+0.3(1+z)^3}/(1+z)^{1.5}$. The emissivity 
derived in this way is consistent with the data of \citet{PoliApJL2001}.
Because of Eqn.~(\ref{normaliz}), the same ratios applies to the 
star-formation histories.

The synthetic spectrum has then been used to derive the emissivity for 
the ionising UV. Due to the absence of observations, synthetic spectra
rely on models of stellar atmospheres for $\lambda\leq 912$\AA.
\citet{CharlotMNRAS2001} compared stellar spectra from different models 
and concluded that discrepancies on the ionising flux are not higher than 
0.1 dex. The uncertainty on our modelled emissivity also depends 
on the uncertainties in the determination of the SFR. To quantify the
uncertainties in the adopted model, we have computed the effect on the 
emissivity of the variation of the basic ingredients of the 
\citet{BruzualPrep} model (IMF, metallicity, stellar libraries).  We
have used the same description for the UV emissivity at 1500\AA\ as in
the main model (solid line in Fig.~\ref{sfrh}) and converted it into 
a star-formation history by using a conversion factor appropriate for
the selected IMF and stellar spectra. The synthetic spectra obtained
in this way typically differ by less than 0.2 dex at the ionisation
limit and 0.3 dex at 600\AA.

Spectra at $\lambda\leq 912$\AA\ also need to be corrected for the 
internal absorption by the galaxy interstellar medium. 
We describe this correction with the parameter $f_\mathrm{esc}$, i.e. 
the fraction of Lyman-continuum photons that can escape into the IGM
without being absorbed by the interstellar medium, either gas or dust.
A wide range of values can be found in the literature for $f_\mathrm{esc}$, 
derived both from models of radiative transfer and observations of \ion{H}{i} 
recombination lines 
\citep[$5\%<f_\mathrm{esc}<60\%$; for a review, see][]{BarkanaPhyRep2001}.
UV observations of local starbursts suggest $f_\mathrm{esc}\approx 5\%$
\citep{LeithererApJL1995,HurwitzApJL1997,HeckmanApJ2001}.
\citet{SteidelApJ2001} analised a composite spectrum of 29 Lyman-break
galaxies at $z\sim 3.4$. They  derived a ratio between the flux
densities at 1500\AA\ and 900\AA\, $f_{1500}/f_{900}=4.6\pm 1$, after
correcting for the differential absorption due to the intervening IGM. 
The $f_{1500}/f_{900}$ ratio for the {\em unattenuated} synthetic spectrum 
that we have used is very similar, $f_{1500}/f_{900}\approx 5.3$.
If we assume that 40\% of the radiation at 1500\AA\ is absorbed by dust
(as obtained from the \citeauthor{CalzettiProc1997}'s attenuation law
with E(B-V)=0.1), the observed $f_{1500}/f_{900}$ ratio is equivalent to
$f_\mathrm{esc}\approx 0.4$ (if the internal absorption in the Lyman
continuum do not change significantly with $\lambda$).
Because of the increase of the disk density with the redshift,
$f_\mathrm{esc}$ is expected to decrease with $z$; it is also found to
depend heavily on the details of the distribution of the sources and the 
gas, i.e. whether the stars and/or gas are clumped or not \citep{WoodApJ2000}.
In this work, we will use a wavelength and redshift independent 
$f_\mathrm{esc}$, by which we multiply the synthetic spectrum at 
$\lambda\leq 912$\AA. We will show results for $f_\mathrm{esc}$ = 0.05 and 
0.40, to cover the range of values suggested by local and $z\sim 3.4$ 
observations, and for an intermediate value, $f_\mathrm{esc}$ = 0.1.

Finally, the galactic emissivity has been converted from comoving to
proper, multiplying by $(1+z)^3$. The total emissivity in
Eqn.~(\ref{transfer}) is the sum of the QSOs and the galaxy
contribution.

\begin{figure*}[ht]
\sidecaption
\includegraphics[width=12cm]{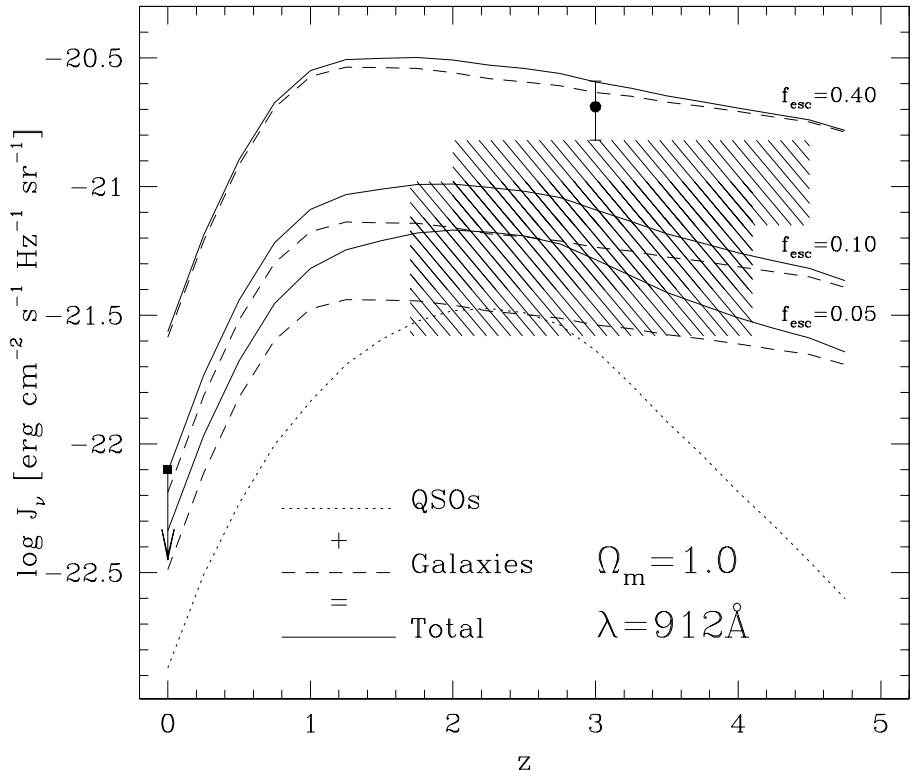}
\caption{UV background at $\lambda=912$\AA\ for the models with
$f_\mathrm{esc}$=0.05, 0.1  and 0.4 (solid lines), in a flat
$\Omega_\mathrm{m}=1$ universe. Also shown are the 
separate contribution of the QSOs (dotted line) and of the galaxies
(dashed lines, each corresponding to a value of $f_\mathrm{esc}$).
The shaded area refer to the Lyman limit UV background estimated from 
the proximity effect \citep{GiallongoApJ1996,CookeMNRAS1997,ScottApJS2000}. 
The arrow shows an upper limit for the local ionising background 
\citep{VogelApJ1995}. The datapoint at $z=3$ is derived from a composite 
spectrum of Lyman-break galaxies \citep{SteidelApJ2001}. The models
and the datapoint of \citet{SteidelApJ2001} have been multiplied by a
$z$-dependent factor, to take into account the cloud emission (see text).
}
\label{j912}
\end{figure*}

\section{The ionising background at 912\AA}
\label{results}

In Fig.~\ref{j912} we show the modelled UV background, $J(\nu,z)$, at the 
Lyman limit as a function of redshift (solid lines) for the flat universe 
with $\Omega_\mathrm{m}$=1. The total background is shown as the sum of 
the QSO contribution (the same in each model; dotted line) and the 
galaxy contribution (scaled with $f_\mathrm{esc}$; dashed lines). As
predicted by other authors, for large values of $f_\mathrm{esc}$ the
ionising background produced by galaxies dominates over the flux from
QSOs \citep{GiallongoMNRAS1997,DevriendtMNRAS1998,ShullAJ1999}. 

At high redshift, the value of the UV background is constrained by the
analysis of the proximity effect, i.e. the decrease in the number of
intervening absorption lines that is observed in a QSO spectrum when 
approaching the QSO's redshift \citep{BajtlikApJ1988}. Using high 
resolution spectra,
\citet{GiallongoApJ1996} derived $J(912$\AA$)=5.0_{-1}^{+2.5} 
\cdot 10^{-22}\;\mathrm{erg\; cm^{-2}\; s^{-1}\; Hz^{-1}\; sr^{-1}}$ for 
$1.7 < z < 4.1$ \citep[see also][]{GiallongoApJ1999}. Larger values are
obtained by \citet*{CookeMNRAS1997}, $J(912$\AA$)=1.0_{-0.3}^{+0.5} \cdot
10^{-21}\;\mathrm{erg\; cm^{-2}\; s^{-1}\; Hz^{-1}\; sr^{-1}}$ for $2.0 <
z < 4.5$.  A recent re-analysis of moderate resolution spectra  by
\citet{ScottApJS2000} has lead to $J(912$\AA$)=7.0_{-4.4}^{+3.4} 
\cdot 10^{-22}\;\mathrm{erg\; cm^{-2}\; s^{-1}\; Hz^{-1}\; sr^{-1}}$ for
the same redshift range. We show these measurements in Fig.~\ref{j912} with
a shaded area: the spread of the measurements obtained with different
methods and data gives an idea of the uncertainties associated with the
study of the proximity effect. 

Measurements of the ionising background at low redshift are not less 
uncertain. \citet{KulkarniApJL1993}
reported a first tentative detection of the proximity effect in a sample
of 13 QSOs at $z<1$ observed with HST. They obtained $J(912$\AA$)=
6_{-4}^{+30} \cdot 10^{-24}\;\mathrm{erg\; cm^{-2}\; s^{-1}\; Hz^{-1}\;
sr^{-1}}$, the large uncertainties
due to the small number of available absorbers. \citet{VogelApJ1995} derived 
a $2\sigma$ upper limit $J$(912\AA)$<8.0 \cdot 10^{-23}\;\mathrm{erg\;
cm^{-2}\; s^{-1}\; Hz^{-1}\; sr^{-1}}$, by studying H$\alpha$ emission in 
a high latitude Galactic cloud. This upper limit is shown in Fig.~\ref{j912}.

We must remember here that our model does not take into account the
Lyman-continuum emission from recombination in Ly$\alpha$ clouds.
\citet{HaardtApJ1996} have shown that radiative recombination provides 
an important contribution to the ionising background. Using only QSOs as
source of ionising radiation, they obtain $J(912$\AA$)=5 \cdot 10^{-22}
\;\mathrm{erg\; cm^{-2}\; s^{-1}\; Hz^{-1}\; sr^{-1}}$ at $z$=2.5, well within
the shaded area in Fig.~\ref{j912}. Similar results are obtained by
\citet{FardalAJ1998}. 
Although the cloud contribution to the background depends on the 
adopted emissivity and intergalactic absorption, we have obtained a rough 
estimate of its importance by using UV background spectra kindly
provided by F. Haardt. At $z$=3, the models of \citet{HaardtApJ1996} are
a factor 1.7 higher than for the case of a purely absorbing medium; at
$z$=0, the factor reduces to 1.3. A simple linear interpolation between
these two points reproduces the actual data for $0<z<5$ within 5\%. 
The models shown in Fig.~\ref{j912} are multiplied for this
$z$-dependent factor.

\begin {figure*}
\centering
\includegraphics[width=17cm]{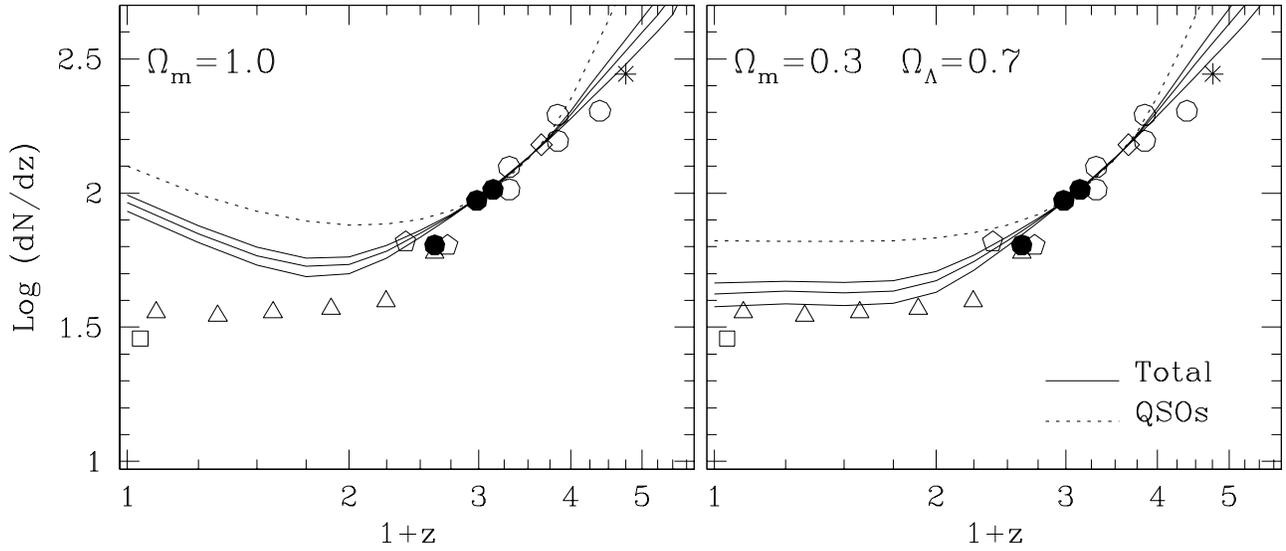}
\caption{Number density evolution of the Ly$\alpha$ forest with 
$N_\ion{H}{i}=10^{13.64-16} \;\mathrm{cm^{-2}}$, for the two cosmologies 
adopted in this paper. Dotted lines refer to the evolution compatible with 
an ionising UV background due only to QSOs. Solid lines show the evolution 
when both QSOs and galaxies contribute to the background, for the models 
with $f_\mathrm{esc}$=0.05 (upper line), 0.1 and 0.4 (lower line). Data points 
come from several observations in the literature for the column density range
$N_\ion{H}{i}=10^{13.64-16}$ cm$^{-2}$, as given by \citet{KimA&A2001}. 
The modelled evolution has been normalized to the observed evolution in 
the redshift range $2<z<3$.}
\label{dndz}
\end{figure*}

Several models can produce a value of $J$(912\AA) compatible with one
of the measurements from the proximity effect at $z\sim 3$ shown in
Fig.~\ref{j912}, from a a simple QSO-dominated background to models with
$f_\mathrm{esc}\sim 0.2$. However, a more stringent condition,
$f_\mathrm{esc}\la 0.10$, is required not to exceed the local upper
limit of \citet{VogelApJ1995}.  \citet{SteidelApJ2001} used their
composite spectrum of Lyman-break galaxies and the UV emissivities of
\citet{SteidelApJ1999} to derive the ionising flux at $z=3$. They
obtained $J(912$\AA$)=1.2\pm0.3\; 10^{-21}\;\mathrm{erg\; cm^{-2}\;
s^{-1}\; Hz^{-1}\; sr^{-1}}$, a value consistent with our model for
$f_\mathrm{esc}=0.4$. If Lyman-break galaxies with a spectrum similar to
that observed by \citet{SteidelApJ2001} dominate the UV background, they
will produce an ionising flux higher than the local and high-redshift
estimates.

Similar results are obtained in the $\Omega_\mathrm{m}$=0.3, 
$\Omega_{\Lambda}=0.7$ universe (not shown). Because of Eqn.~(\ref{dldz}) 
and of the factor we have used in Sect.~\ref{galemi} to derive the 
emissivity for the $\Lambda$-cosmology, the contribution of galaxies to 
the emissivity is exactly the same in both the universe models adopted here. 
The QSOs contribution, instead, depends on an independent fit of the luminosity 
function (Sect.~\ref{qsodis}). In the new cosmology, the UV background 
produced by QSOs is slightly larger than the value for the Einstein-De 
Sitter model, by about 30\% at the peak of the QSOs contribution.~\footnote{As
pointed out by the referee, the background is, in principle, independent
on the adopted cosmology. The slight difference between the two QSO 
backgrounds arises because of the cosmology-dependent correction for 
incompleteness in the luminosity function.}

\section{The evolution of the Lyman Forest}
\label{results2}

Further constraints can be obtained by studying the effect of the
evolution of the ionising background on the evolution of absorbers.
\citet{DaveApJ1999} have studied the evolution of the low-redshift 
Ly$\alpha$ forest in a hydrodynamic cosmological simulation, adopting an UV 
ionising background with the same redshift evolution as that of 
\citet{HaardtApJ1996}. They found a sharp transition at $z=1.7$ for the
number density evolution, $dN/dz$. The change in evolution is primarily
due to the drop in the UV ionising background, resulting from the decline in
the QSO population. The formation of structure by gravitational growth
plays only a minor role in the evolution. In the absence of structure
evolution, is it possible to derive an analytical approximation for
the evolution of $dN/dz$ with $J_\nu$ and the Hubble expansion. For
clouds in photo-ionisation equilibrium with the background, it is
easy to show that the evolution of lines above a given threshold in column
density can be written as \citep{DaveApJ1999}
\begin{equation}
\left(\frac{dN}{dz}\right)_{N_\ion{H}{i}>N_\ion{H}{i}^\mathrm{min}}
\propto 
\frac{1}{H(z)}\;
\left[\frac{(1+z)^5}{\Gamma_\ion{H}{i}(z)}\right]^{\beta-1},
\label{evol}
\end{equation}
where $H(z)$ is the Hubble parameter and $\beta$ the coefficient of the 
power-law distribution of clouds with column density (Eqn.~\ref{absdistr}).
$\Gamma_\ion{H}{i}(z)$ is the photo-ionisation rate
\begin{equation}
\Gamma_\ion{H}{i}(z)=\int_{\nu_0}^{\infty} \frac{4 \pi J(\nu,z)}{h\nu}
\;\sigma_\ion{H}{i}(\nu) d\nu,
\end{equation}
with $\nu_0$ the frequency of the Lyman limit and 
$\sigma_\ion{H}{i}(\nu)=\sigma_\ion{H}{i}(\nu_0/\nu)^3$ the 
\ion{H}{i} photo-ionisation cross-section.

In Fig.~\ref{dndz} we show $dN/dz$ for Lyman forest clouds in the column
density range $N_\ion{H}{i}=10^{13.64-16}\;\mathrm{cm^{-2}}$. Data points 
come from several sources in the literature and from new high resolution 
VLT/UVES spectra of three QSO \citep{KimA&A2001}. For each of our models, 
we have computed $\Gamma_\ion{H}{i}(z)$ and we have derived the evolution of
$dN/dz$ according to Eqn.~(\ref{evol}), for the two cosmologies adopted
in this paper. The evolution has been normalized to the observed values 
for $2<z<3$. In this redshift range, the UV background is nearly flat for 
any of the models and it is easy to show, from Eqn.~(\ref{evol}), that
$dN/dz\propto (1+z)^{5\beta-6.5}$, independently of the cosmology. 
By fitting the observed data, \citet{KimA&A2001} have derived
$dN/dz\propto (1+z)^{2.19}$ for $z>1.5$. A value $\beta\approx 1.7$ can well 
reproduce the evolution derived from the observations. This is consistent 
with fits of the density distribution, that give $\beta= 1.68\pm 0.15$
over $N_\ion{H}{i}=10^{14-16} \;\mathrm{cm^{-2}}$ at $z\sim 1$ 
\citep{KimA&A2001}. On the other hand, weaker lines ($N_\ion{H}{i}\la 
10^{14} \; \mathrm{cm^{-2}}$) are known to have a flatter
distribution in column density \citep[$\beta\sim$ 1.4-1.5; ][]
{GiallongoApJ1996,KimA&A2001}. This will produce a slower redshift 
evolution, as  observed in this column density range for $z>1.5$
\citep[$\gamma\sim$ 1; ][]{KimA&A2001}.
We remind here that in Sect.~\ref{opacity} we have used $\beta=1.46$, that
provides a good description of the density distribution over a much
larger column density range.

The analysis of \citet{KimA&A2001} shows that the change in evolution
occurs at $z\approx 1$, rather than at $z\approx 1.7$, as previously
suggested \citep{WeymannApJ1998}. In Fig.~\ref{dndz} the break at 
$z\approx 1$ can be reproduced if the contribution of galaxies
to the background is dominant. This is because of the rapid decrease 
of the star-formation rate (and of $\Gamma_\ion{H}{i}(z)$) below this
redshift (Fig.~\ref{sfrh}). The photo-ionisation rate of a pure QSOs
background, instead, peaks at $z\sim 2.5$ and has a slower evolution
with $z$. It is interesting to note that the modelled evolution is
closer to the observed data for the $\Lambda$-cosmology. For the
Einstein-De Sitter universe, $dN/dz$ grows for $z<1$, which is not
observed. However, the discrepancy may be mitigated when the effect of the
formation of structures on $dN/dz$ is taken into account \citep{DaveApJ1999}.

The modelled $dN/dz$ do not depend on our approximation of a purely
absorbing intergalactic medium, since the contribution of cloud emission
to $\Gamma_\ion{H}{i}(z)$ is nearly constant with redshift
\citep[see Fig.~6 in][]{HaardtApJ1996}.

\section{Conclusions}
\label{conclu}

In this work we have derived the \ion{H}{i}-ionising background, resulting 
from the integrated contribution of QSOs and galaxies, taking into account 
the opacity of the intervening IGM.
We have modelled the IGM with pure-absorbing clouds, with a distribution
in column density of neutral hydrogen, $N_\ion{H}{i}$, and redshift, z,
derived from recent observations of the Ly$\alpha$ forest
\citep{KimA&A2001} and from Lyman Limit systems. The QSOs emissivity
has been derived from the recent fits of \citet{BoyleMNRAS2000}, while
we have used the stellar population synthesis model of
\citet{BruzualPrep} and a star-formation history from UV observations
for the galaxy emissivity. Due to the present uncertainties in models
and observations, we have used three values for the fraction of ionising
photons that can escape a galaxy interstellar medium, $f_\mathrm{esc}$=
0.05, 0.1 and 0.4, as suggested by local and high-$z$ UV observations of
galaxies, respectively.

The contribution of galaxies to the UV background is found to be comparable 
or larger than that of QSOs. This is consistent with other determinations of 
the galactic contribution to the background
\citep{GiallongoMNRAS1997,DevriendtMNRAS1998,ShullAJ1999}.
Taking into account (in a rough form) the contribution of reemission from 
the IGM clouds \citep{HaardtApJ1996,FardalAJ1998}, we find that all models with
$0<f_\mathrm{esc}\la 0.10$ can provide an ionising background within the
limits measured from observations of the proximity effect at $z\sim 3$
and not exceeding the upper limit for the local background at $z=0$.

The analysis of \citet{KimA&A2001} shows that the break in the evolution
of the Ly$\alpha$ forest, $dN/dz$, occurs at $z\sim 1$. We used the
evolution of the ionising background in our model to derive $dN/dz$,
assuming that the formation of structure plays only a minor role
\citep{DaveApJ1999}. The rapid decrease of star-formation for $z\la 1$
can easily explain the observed break, while a QSO-only background would 
produce a break at $z\sim 2$ \citep{DaveApJ1999}. For the ionising
background to have an evolution similar to that of a galaxy-only model,
high values of $f_\mathrm{esc}$ are needed. This result apparently
pushes the galaxy contribution in an opposite direction with respect to
the estimates of the absolute value of the ionising
background. The flat evolution of $dN/dz$ at $z<1$ is much better
reproduced adopting a $\Lambda$-cosmology 
($\Omega_\mathrm{m}$=0.3,$\Omega_\Lambda$=0.7) rather than a 
$\Omega_\mathrm{m}$=1 Einstein-De Sitter universe. 
However, such a result needs to be confirmed after properly taking the 
formation of structure into account.

A significant contribution from galaxies to the ionising metagalactic flux
would correspond to a softening of its spectrum with respect to a purely
QSO-dominated background. This effect would be particularly important at
very high and very low redshift. Observations of the evolution of the 
\ion{Si}{iv} to \ion{C}{iv} ratio with redshift
\citep{SavaglioA&A1997,SongailaAJ1998} seem to confirm the progressive
softening of the UV background at $z> 3$.

To summarise our result, a galaxy-dominated background with 
$f_\mathrm{esc}\la 0.1$ is consistent with the estimates of  
$J(912\mbox{\AA})$. In the hypothesis that the formation of structures
plays a negligible effect, $f_\mathrm{esc}\ga 0.05-0.1$ is needed in
order to explain the observed $dN/dz$ of Ly$\alpha$ absorbers. 
The rapidly improving knowledge derivable from
numerical simulations, the determination of the cosmological evolution
of the Lyman forest, the proximity effect estimates of the ionising
background, and the evolution of the intensity ratios of metal
absorption lines will put soon constraints on the relative galaxy/QSO
contribution to the UV background and, together with direct measurements
of the Lyman continuum emission from galaxies, will make it possible to
address the issue of the evolution of the escaping fraction of photons
from galaxies as a function of $z$.

\acknowledgements{We are greatly indebted to Francesco Haardt (the
referee) and Piero Madau for providing us with results from their 
cosmological radiative transfer code prior to publication and for 
enlightening comments; and to Stephane Charlot for suggestions and
help with the use of the latest version of the GISSEL code. We also 
acknowledge stimulating discussion with Sandro D'Odorico, Benedetta Ciardi, 
Andrea Ferrara and Andrea Grazian. This work was partially supported by the 
Research and Training Network "The Physics of the Intergalactic Medium" set 
up by the European Community under the contract HPRN-CT2000-00126 RG29185.}

\bibliographystyle{apj}
\bibliography{/home/sbianchi/tex/DUST}

\begin{thebibliography}{52}
\expandafter\ifx\csname natexlab\endcsname\relax\def\natexlab#1{#1}\fi

\bibitem[{{Bajtlik} {et~al.}(1988){Bajtlik}, {Duncan}, \&
  {Ostriker}}]{BajtlikApJ1988}
{Bajtlik}, S., {Duncan}, R.~C., \& {Ostriker}, J.~P. 1988, ApJ, 327, 570

\bibitem[{{Barkana} \& {Loeb}(2001)}]{BarkanaPhyRep2001}
{Barkana}, R. \& {Loeb}, A. 2001, Physics Reports, in press

\bibitem[{{Boyle} {et~al.}(2000){Boyle}, {Shanks}, {Croom}, {Smith}, {Miller},
  {Loaring}, \& {Heymans}}]{BoyleMNRAS2000}
{Boyle}, B.~J., {Shanks}, T., {Croom}, S.~M., {Smith}, R.~J., {Miller}, L.,
  {Loaring}, N., \& {Heymans}, C. 2000, MNRAS, 317, 1014

\bibitem[{{Boyle} {et~al.}(1988){Boyle}, {Shanks}, \&
  {Peterson}}]{BoyleMNRAS1988}
{Boyle}, B.~J., {Shanks}, T., \& {Peterson}, B.~A. 1988, MNRAS, 235, 935

\bibitem[{{Bruzual} \& {Charlot}(1993)}]{BruzualApJ1993}
{Bruzual}, A.~G. \& {Charlot}, S. 1993, ApJ, 405, 538

\bibitem[{{Bruzual} \& {Charlot}(2001)}]{BruzualPrep}
---. 2001, in preparation

\bibitem[{{Calzetti}(1997)}]{CalzettiProc1997}
{Calzetti}, D. 1997, in {AIP} Conference Proceedings, Vol. 408, The Ultraviolet
  Universe at Low and High Redshift, ed. W.~H. e.~a. {Waller} (New York:
  American Institute of Physics), 403

\bibitem[{{Charlot} \& {Longhetti}(2001)}]{CharlotMNRAS2001}
{Charlot}, S. \& {Longhetti}, M. 2001, MNRAS, in press

\bibitem[{{Connolly} {et~al.}(1997){Connolly}, {Szalay}, {Dickinson},
  {Subbarao}, \& {Brunner}}]{ConnollyApJL1997}
{Connolly}, A.~J., {Szalay}, A.~S., {Dickinson}, M., {Subbarao}, M.~U., \&
  {Brunner}, R.~J. 1997, ApJL, 486, L11

\bibitem[{{Cooke} {et~al.}(1997){Cooke}, {Espey}, \&
  {Carswell}}]{CookeMNRAS1997}
{Cooke}, A.~J., {Espey}, B., \& {Carswell}, R.~F. 1997, MNRAS, 284, 552

\bibitem[{{Cristiani} \& {Vio}(1990)}]{CristianiA&A1990}
{Cristiani}, S. \& {Vio}, R. 1990, A\&A, 227, 385

\bibitem[{{Dav{\'e}} {et~al.}(1999){Dav{\'e}}, {Hernquist}, {Katz}, \&
  {Weinberg}}]{DaveApJ1999}
{Dav{\'e}}, R., {Hernquist}, L., {Katz}, N., \& {Weinberg}, D.~H. 1999, ApJ,
  511, 521

\bibitem[{{Devriendt} {et~al.}(1998){Devriendt}, {Sethi}, {Guiderdoni}, \&
  {Nath}}]{DevriendtMNRAS1998}
{Devriendt}, J. E.~G., {Sethi}, S.~K., {Guiderdoni}, B., \& {Nath}, B.~B. 1998,
  MNRAS, 298, 708

\bibitem[{{Fan} {et~al.}(2001){Fan}, {Strauss}, {Schneider}, {Gunn}, {Lupton},
  {Becker}, {Davis}, {Newman}, {Richards}, {White}, {Anderson}, {Annis},
  {Bahcall}, {Brunner}, {Csabai}, {Hennessy}, {Hindsley}, {Fukugita}, {Kunszt},
  {Ivezic}, {Knapp}, {McKay}, {Munn}, {Pier}, {Szalay}, \& {York}}]{FanAJ2001}
{Fan}, X., {Strauss}, M., {Schneider}, D., {Gunn}, J., {Lupton}, R., {Becker},
  R., {Davis}, M., {Newman}, J., {Richards}, G., {White}, R., {Anderson}, J.,
  {Annis}, J., {Bahcall}, N., {Brunner}, R., {Csabai}, I., {Hennessy}, G.,
  {Hindsley}, R., {Fukugita}, M., {Kunszt}, P., {Ivezic}, Z., {Knapp}, G.,
  {McKay}, T., {Munn}, J., {Pier}, J., {Szalay}, A., \& {York}, D. 2001, AJ,
  121, 31

\bibitem[{{Fardal} {et~al.}(1998){Fardal}, {Giroux}, \& {Shull}}]{FardalAJ1998}
{Fardal}, M.~A., {Giroux}, M.~L., \& {Shull}, J.~M. 1998, AJ, 115, 2206

\bibitem[{{Freedman} {et~al.}(2001){Freedman}, {Madore}, {Gibson}, {Ferrarese},
  {Kelson}, {Sakai}, {Mould}, {Kennicutt}, {Ford}, {Graham}, {Huchra},
  {Hughes}, {Illingworth}, {Macri}, \& {Stetson}}]{FreedmanApJ2001}
{Freedman}, W.~L., {Madore}, B.~F., {Gibson}, B.~K., {Ferrarese}, L., {Kelson},
  D.~D., {Sakai}, S., {Mould}, J.~R., {Kennicutt}, R.~C., {Ford}, H.~C.,
  {Graham}, J.~A., {Huchra}, J.~P., {Hughes}, S. M.~G., {Illingworth}, G.~D.,
  {Macri}, L.~M., \& {Stetson}, P.~B. 2001, ApJ, in press

\bibitem[{{Giallongo} {et~al.}(1996){Giallongo}, {Cristiani}, {D'Odorico},
  {Fontana}, \& {Savaglio}}]{GiallongoApJ1996}
{Giallongo}, E., {Cristiani}, S., {D'Odorico}, S., {Fontana}, A., \&
  {Savaglio}, S. 1996, ApJ, 466, 46

\bibitem[{{Giallongo} {et~al.}(1999){Giallongo}, {Fontana}, {Cristiani}, \&
  {D'Odorico}}]{GiallongoApJ1999}
{Giallongo}, E., {Fontana}, A., {Cristiani}, S., \& {D'Odorico}, S. 1999, ApJ,
  510, 605

\bibitem[{{Giallongo} {et~al.}(1997){Giallongo}, {Fontana}, \&
  {Madau}}]{GiallongoMNRAS1997}
{Giallongo}, E., {Fontana}, A., \& {Madau}, P. 1997, MNRAS, 289, 629

\bibitem[{{Haardt} \& {Madau}(1996)}]{HaardtApJ1996}
{Haardt}, F. \& {Madau}, P. 1996, ApJ, 461, 20

\bibitem[{{Heckman} {et~al.}(2001){Heckman}, {Sembach}, {Meurer}, {Leitherer},
  {Calzetti}, \& {Martin}}]{HeckmanApJ2001}
{Heckman}, T., {Sembach}, K.~R., {Meurer}, G., {Leitherer}, C., {Calzetti}, D.,
  \& {Martin}, C.~L. 2001, ApJ, in press

\bibitem[{{Hurwitz} {et~al.}(1997){Hurwitz}, {Jelinsky}, \&
  {Dixon}}]{HurwitzApJL1997}
{Hurwitz}, M., {Jelinsky}, P., \& {Dixon}, W. V.~D. 1997, ApJL, 481, L31

\bibitem[{{Kennicutt}(1998)}]{KennicuttARA&A1998}
{Kennicutt}, R.~C. 1998, ARA\&A, 36, 189

\bibitem[{{Kim} {et~al.}(2001){Kim}, {Cristiani}, \& {D'Odorico}}]{KimA&A2001}
{Kim}, T.~S., {Cristiani}, S., \& {D'Odorico}, S. 2001, A\&A, in press

\bibitem[{{Kulkarni} \& {Fall}(1993)}]{KulkarniApJL1993}
{Kulkarni}, V.~P. \& {Fall}, S.~M. 1993, ApJL, 413, L63

\bibitem[{{La Franca} \& {Cristiani}(1997)}]{LaFrancaAJ1997}
{La Franca}, F. \& {Cristiani}, S. 1997, AJ, 113, 1517

\bibitem[{{Leitherer} {et~al.}(1995){Leitherer}, {Ferguson}, {Heckman}, \&
  {Lowenthal}}]{LeithererApJL1995}
{Leitherer}, C., {Ferguson}, H.~C., {Heckman}, T.~M., \& {Lowenthal}, J.~D.
  1995, ApJL, 454, L19

\bibitem[{{Lilly} {et~al.}(1996){Lilly}, {Le Fevre}, {Hammer}, \&
  {Crampton}}]{LillyApJL1996}
{Lilly}, S.~J., {Le Fevre}, O., {Hammer}, F., \& {Crampton}, D. 1996, ApJL,
  460, L1

\bibitem[{{Liu} {et~al.}(2000){Liu}, {Charlot}, \& {Graham}}]{LiuApJ2000}
{Liu}, M.~C., {Charlot}, S.~., \& {Graham}, J.~R. 2000, ApJ, 543, 644

\bibitem[{{Machacek} {et~al.}(2000){Machacek}, {Bryan}, {Meiksin}, {Anninos},
  {Thayer}, {Norman}, \& {Zhang}}]{MachacekApJ2000}
{Machacek}, M.~E., {Bryan}, G.~L., {Meiksin}, A., {Anninos}, P., {Thayer}, D.,
  {Norman}, M., \& {Zhang}, Y. 2000, ApJ, 532, 118

\bibitem[{{Madau}(1991)}]{MadauApJL1991}
{Madau}, P. 1991, ApJL, 376, L33

\bibitem[{{Madau}(1992)}]{MadauApJL1992}
---. 1992, ApJL, 389, L1

\bibitem[{{Madau} {et~al.}(1996){Madau}, {Ferguson}, {Dickinson}, {Giavalisco},
  {Steidel}, \& {Fruchter}}]{MadauMNRAS1996}
{Madau}, P., {Ferguson}, H.~C., {Dickinson}, M.~E., {Giavalisco}, M.,
  {Steidel}, C.~C., \& {Fruchter}, A. 1996, MNRAS, 283, 1388

\bibitem[{{Madau} {et~al.}(1998){Madau}, {Pozzetti}, \&
  {Dickinson}}]{MadauApJ1998}
{Madau}, P., {Pozzetti}, L., \& {Dickinson}, M. 1998, ApJ, 498, 106

\bibitem[{{Moller} \& {Jakobsen}(1990)}]{MollerA&A1990}
{Moller}, P. \& {Jakobsen}, P. 1990, A\&A, 228, 299

\bibitem[{{Osterbrock}(1989)}]{OsterbrockBook1989}
{Osterbrock}, D.~E. 1989, {Astrophysics of Gaseous Nebulae and Active Galactic
  Nuclei} (Mill Valley: University Science Book)

\bibitem[{{Paresce} {et~al.}(1980){Paresce}, {McKee}, \&
  {Bowyer}}]{ParesceApJ1980}
{Paresce}, F., {McKee}, C.~F., \& {Bowyer}, S. 1980, ApJ, 240, 387

\bibitem[{{Petitjean} {et~al.}(1993){Petitjean}, {Webb}, {Rauch}, {Carswell},
  \& {Lanzetta}}]{PetitjeanMNRAS1993}
{Petitjean}, P., {Webb}, J.~K., {Rauch}, M., {Carswell}, R.~F., \& {Lanzetta},
  K. 1993, MNRAS, 262, 499

\bibitem[{{Poli} {et~al.}(2001){Poli}, {Menci}, {Giallongo}, {Fontana},
  {Cristiani}, \& {D'Odorico}}]{PoliApJL2001}
{Poli}, F., {Menci}, N., {Giallongo}, E., {Fontana}, A., {Cristiani}, S., \&
  {D'Odorico}, S. 2001, ApJL, in press

\bibitem[{{Salpeter}(1955)}]{SalpeterApJ1955}
{Salpeter}, E.~E. 1955, ApJ, 121, 161

\bibitem[{{Savaglio} {et~al.}(1997){Savaglio}, {Cristiani}, {D'Odorico},
  {Fontana}, {Giallongo}, \& {Molaro}}]{SavaglioA&A1997}
{Savaglio}, S., {Cristiani}, S., {D'Odorico}, S., {Fontana}, A., {Giallongo},
  E., \& {Molaro}, P. 1997, A\&A, 318, 347

\bibitem[{{Schechter}(1976)}]{SchechterApJ1976}
{Schechter}, P. 1976, ApJ, 203, 297

\bibitem[{{Scott} {et~al.}(2000){Scott}, {Bechtold}, {Dobrzycki}, \&
  {Kulkarni}}]{ScottApJS2000}
{Scott}, J., {Bechtold}, J., {Dobrzycki}, A., \& {Kulkarni}, V.~P. 2000, ApJS,
  130, 67

\bibitem[{{Shull} {et~al.}(1999){Shull}, {Roberts}, {Giroux}, {Penton}, \&
  {Fardal}}]{ShullAJ1999}
{Shull}, J.~M., {Roberts}, D., {Giroux}, M.~L., {Penton}, S.~V., \& {Fardal},
  M.~A. 1999, AJ, 118, 1450

\bibitem[{{Songaila}(1998)}]{SongailaAJ1998}
{Songaila}, A. 1998, AJ, 115, 2184

\bibitem[{{Steidel} {et~al.}(1999){Steidel}, {Adelberger}, {Giavalisco},
  {Dickinson}, \& {Pettini}}]{SteidelApJ1999}
{Steidel}, C.~C., {Adelberger}, K.~L., {Giavalisco}, M., {Dickinson}, M., \&
  {Pettini}, M. 1999, ApJ, 519, 1

\bibitem[{{Steidel} {et~al.}(2001){Steidel}, {Pettini}, \&
  {Adelberger}}]{SteidelApJ2001}
{Steidel}, C.~C., {Pettini}, M., \& {Adelberger}, K.~L. 2001, ApJ, 546, 665

\bibitem[{{Storrie-Lombardi} {et~al.}(1994){Storrie-Lombardi}, {McMahon},
  {Irwin}, \& {Hazard}}]{StorrieLombardiApJL1994}
{Storrie-Lombardi}, L.~J., {McMahon}, R.~G., {Irwin}, M.~J., \& {Hazard}, C.
  1994, ApJL, 427, L13

\bibitem[{{Vogel} {et~al.}(1995){Vogel}, {Weymann}, {Rauch}, \&
  {Hamilton}}]{VogelApJ1995}
{Vogel}, S.~N., {Weymann}, R., {Rauch}, M., \& {Hamilton}, T. 1995, ApJ, 441,
  162

\bibitem[{{Weymann} {et~al.}(1998){Weymann}, {Jannuzi}, {Lu}, {Bahcall},
  {Bergeron}, {Boksenberg}, {Hartig}, {Kirhakos}, {Sargent}, {Savage},
  {Schneider}, {Turnshek}, \& {Wolfe}}]{WeymannApJ1998}
{Weymann}, R.~J., {Jannuzi}, B.~T., {Lu}, L., {Bahcall}, J.~N., {Bergeron}, J.,
  {Boksenberg}, A., {Hartig}, G.~F., {Kirhakos}, S., {Sargent}, W. L.~W.,
  {Savage}, B.~D., {Schneider}, D.~P., {Turnshek}, D.~A., \& {Wolfe}, A.~M.
  1998, ApJ, 506, 1

\bibitem[{{Wood} \& {Loeb}(2000)}]{WoodApJ2000}
{Wood}, K. \& {Loeb}, A. 2000, ApJ, 545, 86

\bibitem[{{Zheng} {et~al.}(1997){Zheng}, {Kriss}, {Telfer}, {Grimes}, \&
  {Davidsen}}]{ZhengApJ1997}
{Zheng}, W., {Kriss}, G.~A., {Telfer}, R.~C., {Grimes}, J.~P., \& {Davidsen},
  A.~F. 1997, ApJ, 475, 469

\end{thebibliography}

\end{document}